# Theoretical evidence for the superluminality of evanescent modes

**Zhi-Yong Wang**, **Cai-Dong Xiong**

*School of Physical Electronics, University of Electronic Science and Technology of China, Chengdu 610054, CHINA*

Though both theoretical and experimental investigations have revealed the superluminal behavior of evanescent electromagnetic waves, there are many disputes about the physical meaning and validity of such superluminal phenomenon, which is due to the fact that the traditional investigations are based on the theory of tunneling time, and concerned with the problem of what the group velocity of evanescent waves means. In this paper, by studying the quantum probability amplitude for photons to propagate over a spacelike interval along an undersized waveguide, we present theoretical evidence for such superluminality.

PACS number(s): 03.65.Xp, 41.20.Jb, 42.50.Nn, 03.65.Pm

**I. INTRODUCTION**

From the point of view of classical electromagnetic field theory, inside an undersized waveguide evanescent waves have support everywhere (through exponential damping) along the undersized waveguide, and "the propagation of evanescent modes" is not a well-defined concept; but, from the quantum-mechanical viewpoint, there exists a physical process that photons propagate through an undersized waveguide (i.e., the so-called photon tunneling phenomenon), and then we can study the propagation of evanescent modes in the sense of quantum mechanics. To do so, traditionally, people have applied a formal analogy between the Helmholtz equation describing evanescent modes and the nonrelativistic Schrödinger equation for a particle tunneling through a potential barrier [1-2], and by which the theory of tunneling time is applied to reveal the superluminal behavior of evanescent waves. However, such an analogy is appropriate in mathematics, not in physics, because the photon's equation of motion is relativistic while the Schrödinger equation is not. Moreover, time in quantum mechanics has been a controversial issue since the advent of quantum theory, and there are a lot of theoretical models of tunneling time [3-6]. As a consequence, in spite of the fact that theoretical and experimental studies had obtained the same conclusion that photons inside an undersized waveguide propagate superluminally [7-13], many papers disproving this conclusion have been published recently [14-20], they are mainly based on reinterpreting the physical meaning of the evanescent waves' group velocity.

A proper description for the propagation of evanescent modes should be based on the photon's quantum theory itself rather than just via a quantum-mechanical analogy. Therefore, in this paper we will try to study the propagation of photons inside an undersized waveguide at the level of quantum field theory of photons. That is, by studying the quantum probability amplitude for photons to propagate along an undersized waveguide, we present theoretical evidence for the superluminal behavior of evanescent modes. What we study is the superluminality of evanescent modes (i.e., to show that photons' motion through an undersized waveguide is a spacelike one), and have nothing to do with the controversy about how to interpret the physical meaning of the evanescent waves' group velocity. This paper does not study whether a signal (information) can propagate superluminally, this issue can be found in Refs. [21, 22], where QED-based studies of evanescent modes are given, and a successful test of these predictions with experimental data has been presented.

In the following, the natural units of measurement ($\hbar = c = 1$) is applied, repeated indices must be summed according to the Einstein rule, and the space-time metric tensor is chosen as $g^{\mu\nu} = \text{diag}(1,-1,-1,-1)$, $\mu, \nu = 0,1,2,3$, $x = x^\mu = (t, \boldsymbol{x})$, and so on.

## II. SPACElike BEHAVIORS OF PHOTONS INSIDE AN UNDERSIZED WAVEGUIDE

Let $\varphi(x)$ represent a field operator and $|0\rangle$ denote the field's vacuum state. According to quantum field theory, the quantity $D(x-y) \equiv \langle 0|\varphi(x)\varphi(y)|0\rangle$ represents a transition probability amplitude from the quantum state $\varphi(y)|0\rangle$ to the quantum state $\varphi(x)|0\rangle$, such that $|D(x-y)|^2$ is related to the probability for a particle to propagate from $y^\mu = (y^0, \boldsymbol{y})$ to $x^\mu = (x^0, \boldsymbol{x})$. If $\varphi(x)$ is the Klein-Gordon field, then $D(x-y)$ represents the probability amplitude for a scalar particle to propagate from $y$ to $x$ [23]. In particular, if its mass vanishes ($m=0$), one can obtain (see, for example, Ref. [23] with the replacements $p^\mu = (p^0, \boldsymbol{p}) \to k^\mu = (k^0, \boldsymbol{k})$ and $E_p = \sqrt{\boldsymbol{p}^2 + m^2} \to \omega = \sqrt{\boldsymbol{k}^2}$)

$$D(x-y) = \langle 0|\varphi(x)\varphi(y)|0\rangle = \int \frac{d^3 k}{(2\pi)^3} \frac{1}{2\omega} \exp[-ik \cdot (x-y)], \qquad (1)$$

where $k \cdot (x-y) = k^\mu (x_\mu - y_\mu)$ (and so on), $k^\mu = (k^0, \boldsymbol{k})$ is the four-dimensional (4D) momentum of the scalar particle and $\omega = \sqrt{\boldsymbol{k}^2}$. Now, let $k^\mu = (k^0, \boldsymbol{k})$ denote the 4D momentum of photons in free vacuum, $|0\rangle$ denote the photons' vacuum state, and $A^\mu(x)$ denote the 4D electromagnetic potential. In quantum field theory, $A^\mu(x)$ plays the role of the field operator. In Lorentz gauge condition, it is well known that $A^\mu(x)$ satisfies ($\mu, \nu = 0,1,2,3$):

$$R^{\mu\nu}(x-y) \equiv \langle 0|A^\mu(x) A^\nu(y)|0\rangle = -g^{\mu\nu} D(x-y), \qquad (2)$$

where $g^{\mu\nu} = \text{diag}(1,-1,-1,-1)$ is the space-time metric tensor and $D(x-y)$ is given by Eq. (1) with $k^\mu = (k^0, \boldsymbol{k})$ being reinterpreted as the 4D momentum of photons.

In a Cartesian coordinate system spanned by an orthonormal basis $\{\boldsymbol{e}_1, \boldsymbol{e}_2, \boldsymbol{e}_3\}$ with $\boldsymbol{e}_3 = \boldsymbol{e}_1 \times \boldsymbol{e}_2$, we assume that a hollow metallic waveguide is placed along the direction of $\boldsymbol{e}_3$, and the waveguide is a straight rectangular pipe with the cross-sectional dimensions $b_1$ and $b_2$, let $b_1 < b_2$ without loss of generality. Then the cutoff frequency of the waveguide is $\omega_{crs} = \pi\sqrt{(r/b_1)^2 + (s/b_2)^2}$ ($r = 0, 1, 2, ...$, $s = 1, 2, 3...$). For simplicity, we shall restrict our discussion to the lowest-order cutoff frequency $\omega_c = \pi/b_2$. It is also assumed that the waveguide is infinitely long and its conductivity is infinite, and the electromagnetic source is localized at infinity. In the present case, as for the 4D momentum $k^\mu = (k^0, \boldsymbol{k})$ of photons in the free space inside the waveguide, its first the second components are fixed: $k_1 = 0$ and $k_2 = \pi/b_2$, such that the function $D(x-y)$ presented in Eq. (2) becomes:

$$\begin{aligned}D(x-y) &= \int \frac{\mathrm{d}^3 k}{(2\pi)^3} \delta(k_1) \delta(k_2 - \pi/b_2) \frac{1}{2\omega} \exp[-\mathrm{i}k \cdot (x-y)] \\ &= P \int \frac{\mathrm{d}k_3}{2\pi} \frac{1}{2\omega} \exp[-\mathrm{i}\omega(x_0 - y_0) + \mathrm{i}k_3(x_3 - y_3)]\end{aligned} \qquad (3)$$

where $\omega = k^0 = \sqrt{\omega_c^2 + k_3^2}$ is the frequency of photons and $P = \exp[\mathrm{i}(\pi/b_2)(x_2 - y_2)]/(2\pi)^2$ is a phase factor. Because $A^\mu(x)$ is not an observable, to study the superluminal behavior of photons inside an undersized waveguide in a strict manner, our discussion will be based on field intensities (they are observables). For example, in terms of the electric field intensity $E^i = \partial^i A^0 - \partial^0 A^i$ (it is an operator in the second-quantization sense), one can analyze $S_{ij}(x-y) \equiv \langle 0 | E_i(x) E_j(y) | 0 \rangle$ without loss of generality ($i, j = 1, 2, 3$). Likewise, the function $S_{ij}(x-y)$ is related to the probability amplitude for photons to propagate from $y$ to $x$ along the waveguide, it can also be regarded as a correlation function for the electromagnetic field. Therefore, in the sense of quantum field theory, one can study the propagation of evanescent modes via the function $S_{ij}(x-y)$. If $S_{ij}(x-y) \neq 0$ for a spacelike interval $(x-y)$ along the undersized waveguide, then evanescent modes have superluminal behavior. Substituting $E^i = \partial^i A^0 - \partial^0 A^i$ into $S_{ij}(x-y)$, one can obtain

$$S_{ij}(x-y) \equiv \langle 0|E_i(x)E_j(y)|0\rangle = (\frac{\partial^2}{\partial x^i \partial x^j} - \delta_{ij}\frac{\partial^2}{\partial x_0^2})D(x-y). \tag{4}$$

For our purpose, we will only consider the component $S_{11}(x-y)$ without loss of generality. As for photons inside the undersized waveguide, one has $0 < \omega = \sqrt{\omega_c^2 + k_3^2} < \omega_c$, which implies that $k_3 = iq$, where $-\omega_c < q < \omega_c$. Because the waveguide is placed along the third axes, one can take $y = (0,0,0,0)$ and $x = (t,0,0,r)$ (for simplicity let $t, r \geq 0$), and then in Eq. (3) the factor $P = 1/(2\pi)^2$. Moreover, Eq. (3) shows that the function $D(x-y)$ is independent of $x_1$ and $y_1$, such that $\partial D/\partial x_1 = 0$. Using $\partial D/\partial x_1 = 0$ and taking $y = (0,0,0,0)$, $x = (t,0,0,r)$, the function $S_{11}(x-y) = S_{11}(x) = S_{11}(t,r)$ becomes:

$$S_{11}(x) = (\frac{\partial^2}{\partial x_1^2} - \frac{\partial^2}{\partial t^2})D(x) = -\frac{\partial^2}{\partial t^2}D(x), \tag{5}$$

By substituting $y = (0,0,0,0)$, $x = (t,0,0,r)$, $k_3 = iq$ and $0 < \omega = \sqrt{\omega_c^2 + k_3^2} < \omega_c$ into Eq. (3), one can obtain [for convenience the constant factor $P = 1/(2\pi)^2$ is omitted]

$$D(x) = \int_0^{\omega_c} \frac{dq}{2\pi} \frac{1}{2\sqrt{\omega_c^2 - q^2}} \exp(-i\sqrt{\omega_c^2 - q^2}\, t - qr). \tag{6}$$

Here, it is necessary to point out that, if photons propagating along an undersized waveguide from A to B correspond to evanescent waves, then photons propagating along the undersized waveguide from B to A would correspond to antievanescent waves. Obviously, along the direction of A→B, the amplitudes of the evanescent waves are exponential damping, while those of the antievanescent waves are exponential increasing.

Because our study is focused on the probability amplitude for photons to propagate from $y = (0,0,0,0)$ to $x = (t,0,0,r)$, Eq. (6) does not contain the contribution of the anti-evanescent waves propagating from $x = (t,0,0,r)$ to $y = (0,0,0,0)$, and then the integrating range in Eq. (6) is taken as $(0, \omega_c)$ rather than $(-\omega_c, \omega_c)$.

Note that $\omega_c = \pi/b_2$ is the lowest-order cutoff frequency, and Eqs. (1)-(4) are written in an arbitrary inertial frame of reference.

In order to evaluate the integral Eq. (6), for timelike interval $x^2 = x_\mu x^\mu = t^2 - r^2 > 0$, let $t = \sqrt{x^2}\cosh\phi$ and $r = \sqrt{x^2}\sinh\phi$, and there is always an inertial frame in which $r = 0$; for spacelike interval $x^2 < 0$, let $t = \sqrt{-x^2}\sinh\phi$ and $r = \sqrt{-x^2}\cosh\phi$, and there is always an inertial frame in which $t = 0$. As $\phi$ varies in $[0, +\infty)$, $x^2$ is Lorentz invariant, then for convenience we will take $\phi \to 0$. Furthermore, the integral representation of the Hankel function of the second kind is useful:

$$H_0^{(2)}(z) = \frac{2}{\pi} \int_0^{\pi/2} d\theta \exp(-iz\sin\theta). \tag{7}$$

The Hankel function behaves for large arguments $|z|$ as

$$H_\nu^{(2)}(z) \sim \sqrt{2/\pi z}\exp[-i(z - \pi\nu/2 - \pi/4)], \quad |z| \to +\infty. \tag{8}$$

From Eq. (6) one can obtain

$$D(x) = \begin{cases} \dfrac{1}{8} H_0^{(2)}(\omega_c \sqrt{x^2}) & \text{for timelike interval } x^2 > 0, \\ \dfrac{1}{8} H_0^{(2)}(-i\omega_c \sqrt{-x^2}) & \text{for spacelike interval } x^2 < 0. \end{cases} \tag{9}$$

The Hankel functions satisfy the recurrence relations $z^{-\nu} H_{\nu+1}^{(2)}(z) = -d[z^{-\nu} H_\nu^{(2)}(z)]/dz$. Applying Eqs. (5) and (9), consider that the relations $H_1^{(2)}(z) = -d[H_0^{(2)}(z)]/dz$ and $z^{-1} H_2^{(2)}(z) = -d[z^{-1} H_1^{(2)}(z)]/dz$, where $z = \omega_c \sqrt{t^2 - r^2}$, one can obtain:

$$S_{11}(x) = \begin{cases} \dfrac{\omega_c}{8\sqrt{x^2}}[H_1^{(2)}(\omega_c\sqrt{x^2}) - tH_2^{(2)}(\omega_c\sqrt{x^2})], & \text{for } x^2 > 0, \\ \dfrac{i\omega_c}{8\sqrt{-x^2}}[H_1^{(2)}(-i\omega_c\sqrt{-x^2}) - tH_2^{(2)}(-i\omega_c\sqrt{-x^2})], & \text{for } x^2 < 0. \end{cases} \quad (10)$$

Equation (8) implies that for large timelike interval ($x^2 \to +\infty$) one has $S_{11}(x) \sim \exp(-i\omega_c\sqrt{x^2})$; for large spacelike interval ($x^2 \to -\infty$) one has $S_{11}(x) \sim \exp(-\omega_c\sqrt{-x^2})$. To be specific, consider the fact that for timelike interval $x^2 = t^2 - r^2 > 0$, there is always an inertial frame in which $r = 0$; for spacelike interval $x^2 = t^2 - r^2 < 0$, there is always an inertial frame in which $t = 0$, one can deduce the asymptotic behaviors of $S_{11}(x)$ as follows:

$$S_{11}(x) \sim \begin{cases} (t)^{-1/2}\exp(-i\omega_c t), & \text{as timelike interval } x^2 = t^2 \to +\infty, \\ (r)^{-3/2}\exp(-\omega_c r), & \text{as spacelike interval } x^2 = -r^2 \to -\infty. \end{cases} \quad (11)$$

It is very important to note that, because the evanescent waves oscillate with time as $\exp(-i\omega t)$, as observed in an inertial frame of reference moving relative to the waveguide, the evanescent waves can propagate with a velocity $v < c = 1$, which is related to the Lorentz transformation of $\omega t$ presented in the factor of $\exp(-i\omega t)$. Therefore, though $S_{11}(x) \neq 0$ for timelike interval $x^2 > 0$, its physical meaning is trivial. On the other hand, the propagation of evanescent waves through the waveguide is characterized by an exponential damping factor, and Eqs. (10)-(11) show that such a damping propagation is actually a spacelike (i.e., superluminal) one.

### III. CONCLUSIONS AND DISCUSSIONS

As discussed before, according to quantum field theory, the function $S_{11}(x)$ given by Eq. (5) is related to the probability amplitude for photons to propagate from $y = (0,0,0,0)$ to $x = (t,0,0,r)$ along the undersized waveguide. Using Eqs. (10) and (11) one can show that for timelike intervals ($x^2 = t^2 - r^2 > 0$), the function $S_{11}(x)$ are the oscillating ones slowly decreasing in amplitude owing to the power-law factor, this behavior is just related to the fact that the evanescent waves contain the factor of $\exp(-i\omega t)$ and its physical meaning is trivial. For spacelike intervals ($x^2 = t^2 - r^2 < 0$), the function $S_{11}(x)$ rapidly fall to zero according to the exponential function (with the scale being set by the inverse cutoff frequency of the waveguide), this behavior corresponds to the propagation of photons through the undersized waveguide. In other words, the propagation of photons along the undersized waveguide is superluminal.

The superluminal behavior of photons through an undersized waveguide is due to a purely quantum-mechanical effect, and it preserves weak causality that has been discussed in Refs. [24-27], where in Refs. [24-26] weak causality is discussed on a single particle level, while in Ref. [27] weak causality is discussed for ensembles of particles in field theory. In our case, the superluminal behavior preserves Einstein causality for expectation values or ensemble average only, not for individual process (in the sense of which weak causality can also be called quantum-mechanical causality). Within local quantum field theory a rigorous proof of weak causality for local observables has been given in the previous literatures [27]. To avoid a possible causality paradox, one can also resort to the particle-antiparticle symmetry. The process of a particle created at *x* and annihilated at *y* as observed in a frame of reference is identical with that of an antiparticle created at *y* and annihilated at *x* as observed in another frame of reference [28]. In our case, the antiparticle of the photon is the photon itself. Therefore, the process that a photon propagates superluminally from A to B as observed in a frame of reference is equivalent to one where the photon propagates superluminally from B to A as observed in another frame of reference, where causality is preserved provided that every observer has a consistent causal history locally. As we know, in total internal reflection, photons lie in evanescent modes and are actually virtual photons that describe the excitations of coupled modes of photons with matter [29, 30]. These virtual photons are the carrier of electromagnetic interaction in the quantum regime, and lie outside the ordinary photon dispersion relation and cannot be observed outside of the interacting system, but can be observed inside the system by means of a destructive measurement via a direct interaction with a probe [30]. On the other hand, Feynman has presented another way of looking at the guided waves [31], by which one can show that inside an undersized waveguide photons are also virtual photons.

**ACKNOWLEDGMENTS**

The first author (Z. Y. W.) would like to thank Professor G. Nimtz for helpful discussions. This work was supported by the Specialized Research Fund for the Doctoral Program of Higher Education of China (Grant No. 20050614022) and by the National Natural Science Foundation of China (Grant No. 60671030).

New York, 1964), Vol. 2, Chap. 24-8.